\documentclass[11pt]{article}

\usepackage{latexsym}
\usepackage{amssymb}
\usepackage{framed}
\usepackage{multirow}
\usepackage{bbm}
\def\thefootnote{\fnsymbol{footnote}}

\DeclareMathAlphabet   {\mathsc}{OT1}{cmr}{m}{sc}

\newcommand{\myref}[1]{(\ref{#1})}

\newcommand{\myapp}[1]{Appendix~\ref{#1}}

\def\Del{\Delta}
\def\half{{1\over2}}
\def\bea{\begin{eqnarray}}
\def\eea{\end{eqnarray}}
\def\beq{\begin{equation}}
\def\eeq{\end{equation}}

\def\uo{$U(1)$}

\def\superint{\int d^{4}\theta}

\newcommand{\WaWa}{ W_a^{\alpha} W^a_{\alpha}}

\newcommand{\DaDa}{{\cal D}^{\alpha}{\cal D}_{\alpha}}
\newcommand{\DbDb}{{\cal D}_{\dot{\alpha}}{\cal D}^{\dot{\alpha}}}
\newcommand{\Da}{{\cal D}_{\alpha}}
\newcommand{\Db}{{\cal D}^{\dot{\beta}}}

\newcommand{\Dc}{{\cal D}^{\alpha}}
\newcommand{\Dd}{{\cal D}_{\dot{\beta}}}
\newcommand{\Wa}{W_{\alpha}}

\newcommand{\Xa}{X_{\alpha}}

\newcommand{\Ga}{G_{\alpha}}
\newcommand{\Xc}{X^{\alpha}}
\newcommand{\La}{L_{\alpha}}
\newcommand{\Lc}{L^{\alpha}}
\newcommand{\fa}{f_{\alpha}}
\newcommand{\fc}{f^{\alpha}}
\newcommand{\ga}{g_{\alpha}}
\newcommand{\gc}{g^{\alpha}}

\def\vev{$vev$}
\def\re{{\rm Re}}
\def\im{{\rm Im}}

\def\tg{\tilde{g}}

\def\G{{\cal G}}

\def\tG{{\widetilde G}}
\def\tX{{\widetilde X}}

\def\tF{{\tilde F}}

\def\dZ{{\dot Z}}
\def\dY{{\dot Y}}

\def\D{{\cal D}}

\def\bD{\bar{\D}}

\def\pp{\partial}

\def\ibar{\bar{\imath}}

\def\[{\left [}
\def\]{\right ]}
\def\({\left (}
\def\){\right )}
\def\lbr{\left\{}
\def\rbr{\right\}}
\def\r{\right|}
\def\l{\left.}

\def\T{\bar{T}}

\def\z{\bar{z}}
\def\q{\bar{q}}
\def\del{\delta}

\def\deta{{\dot\eta}}

\def\S{{\bar{S}}}

\def\Tr{{\rm Tr}}

\def\Ti^{T^{(i)}}

\def\f{\bar{f}}
\def\bR{\bar{R}}

\def\L{{\cal L}}

\def\t{\bar{t}}

\def\n{\bar{n}}
\def\m{\bar{m}}
\def\s{\bar{s}}
\def\cM{{\cal{M}}}

\def\Z{{\bar{Z}}}

\def\bF{\bar{F}}

\def\a0{\alpha_0}
\def\chiproj{(\bD^2 - 8R)}

\def\eee{\nonumber \\ &=&}
\def\ddd{\nonumber \\ &&}
\def\mmm{\nonumber \\ &&\qquad}
\def\nnn{\nonumber \\ }
\def\hc{ + {\rm h.c.}}

\begin{document}

\begin{titlepage}
\begin{center}

\hfill \today \\[1in]

{\large {\bf ANOMALY CANCELLATION IN EFFECTIVE SUPERGRAVITY THEORIES
    FROM THE HETEROTIC STRING: TWO SIMPLE EXAMPLES}}\footnote{This
  work was supported in part by the Director, Office of Science,
  Office of High Energy and Nuclear Physics, Division of High Energy
  Physics, of the U.S. Department of Energy under Contract
  DE-AC02-05CH11231 and in part by the National
  Science Foundation under grants PHY-1316783.}  \\[.1in]

 Mary K. Gaillard and Jacob Leedom\\[.1in]

{\em Department of Physics and Theoretical Physics Group,
 Lawrence Berkeley Laboratory, \\
 University of California, Berkeley, California 94720}\\[1in] 

\end{center}

\begin{abstract}

We use Pauli-Villars regularization to evaluate the conformal and
chiral anomalies in the effective field theories from $Z_3$ and $Z_7$
compactifications of the heterotic string without Wilson lines.  We
show that parameters for Pauli-Villars chiral multiplets can be chosen
in such a way that the anomaly is universal in the sense that its
coefficient depends only on a single holomorphic function of the three
diagonal moduli.  It is therefore possible to cancel the anomaly by a
generalization of the four-dimensional Green-Schwarz mechanism.  In
particular we are able to reproduce the results of a string
calculation of the four-dimensional chiral anomaly for these two models.
  
\end{abstract}
\end{titlepage}

\newpage

\renewcommand{\thepage}{\roman{page}}
\setcounter{page}{2}
\mbox{ }

\begin{center}
{\bf Disclaimer}
\end{center}

\vskip .2in

\begin{scriptsize}
\begin{quotation}
  This document was prepared as an account of work sponsored by the
  United States Government. While this document is believed to contain
  correct information, neither the United States Government nor any
  agency thereof, nor The Regents of the University of California, nor
  any of their employees, makes any warranty, express or implied, or
  assumes any legal liability or responsibility for the accuracy,
  completeness, or usefulness of any information, apparatus, product,
  or process disclosed, or represents that its use would not infringe
  privately owned rights.  Reference herein to any specific commercial
  products process, or service by its trade name, trademark,
  manufacturer, or otherwise, does not necessarily constitute or imply
  its endorsement, recommendation, or favoring by the United States
  Government or any agency thereof, or The Regents of the University
  of California.  The views and opinions of authors expressed herein
  do not necessarily state or reflect those of the United States
  Government or any agency thereof, or The Regents of the University
  of California.
\end{quotation}
\end{scriptsize}

\vskip 2in

\begin{center}
\begin{small}
{\it Lawrence Berkeley Laboratory is an equal opportunity employer.}
\end{small}
\end{center}

\newpage
\renewcommand{\theequation}{\arabic{section}.\arabic{equation}}
\renewcommand{\thepage}{\arabic{page}}
\setcounter{page}{1}
\def\thefootnote{\arabic{footnote}}
\setcounter{footnote}{0}

\section{Introduction}
On-shell Pauli-Villars regularization of the one-loop divergences of
supergravity theories was used to determine the anomaly structure of
supergravity in~\cite{bg}.  Pauli-Villars regulator fields allow for
the cancellation of all quadratic and logarithmic
divergences~\cite{pv}, as well as most linear
divergences~\cite{bg}. If all linear divergences were canceled, the
theory would be anomaly free, with noninvariance of the action arising
only from Pauli-Villars masses.  However there are linear divergences
associated with nonrenormalizable gravitino/gaugino interactions that
cannot be canceled by PV fields.  The resulting chiral anomaly forms a
supermultiplet with the corresponding conformal anomaly, provided the
ultraviolet cut-off has the appropriate field dependence, in which
case uncanceled total derivative terms, such as Gauss-Bonnet, do not
drop out from the effective action.  The resulting anomaly term that
is quadratic in the field strength associated with the space-time
curvature, as well as the term quadratic in the Yang-Mills field
strength, was shown in~\cite{bg} to be canceled by the
four-dimensional version of the Green-Schwarz mechanism in $Z_3$ and
$Z_7$ compactifications, in agreement with earlier
results~\cite{prev}.  However, the terms in the anomaly that are
quadratic and cubic in the parameters of the anomalous transformation
are prescription dependent~\cite{ss,bg}.  The choice of PV fields with
noninvariant masses used in~\cite{bg} did not achieve full anomaly
cancellation.

Every contribution to the chiral anomaly has a conformal anomaly
counterpart, with which it combines to form an ``F-term'' anomaly.
In addition there are ``D-term'' anomalies associated with logarithmic
divergences that have no chiral partner.  In a generic supergravity
theory, these include terms~\cite{bg} that are nonlinear in the 
holomorphic
functions $F^i(T^i)$ of the three diagonal K\"ahler moduli $T^i$ that
characterize modular (or T-duality) transformations:
\bea T'^i &=& {a_i - ib_i T^i\over i c_i T^i + d_i}, \qquad a_i b_i 
- c_i d_i = 1,
\qquad a_i,b_i,c_i,d_i \in {\bf Z},\nnn
\Phi'^a &=& e^{-\sum_i q^a_i F^i(T^i)}\Phi^a,\qquad F^i(T^i) = 
\ln(i c_i T^i + d_i),\label{modtr}\eea
where $\Phi^a$ is any chiral supermultiplet other than a diagonal
K\"ahler modulus, and $q^a_i$ are its modular weights.  Only terms in
the anomaly that are linear in $F = \sum_i F^i$ can be canceled by
the Green-Schwarz term.

In addition, in generic supergravity there are anomalous terms that
involve the dilaton superfield $S$ in the chiral supermultiplet
formulation--or $L$ in the linear multiplet formulation~\cite{linear}
for the dilaton.  Specifically, one expects~\cite{kahler} a term
quadratic in the K\"ahler field strength
\beq X_{\mu\nu} = \(\D_\mu z^i\D_\nu\z^{\m} - 
\D_\nu z^i\D_\mu\z^{\m}\)K_{i\m} - iF^a_{\mu\nu}(T_a z^i)K_i,
\label{xmunu}\eeq
\noindent 
where $z^i = \l Z^i\r$ is the scalar component of a generic chiral
superfield $Z^i$, $F^a_{\mu\nu}$ is the gauge field strength, $T_a$ is
a gauge group generator, and $K(Z,\Z)$ is the K\"ahler potential.  The
term quadratic in $X_{\mu\nu}$ was actually found to vanish
in~\cite{bg}, but there remained terms linear in $X_{\mu\nu}$ as
well as terms involving the K\"ahler potential in the nonlinear $F^i$
terms mentioned above.  Anomaly cancellation by a Green-Schwarz
mechanism, to be outlined in the next section, requires that the
operators appearing in the anomaly also appear in the real superfield
$\Omega$ of the (modified) linearity condition for the superfield $L$:
\beq \(\bD^2 - 8R\)\(L + \Omega\) = \(\D^2 - 8\bR\)\(L + \Omega\) = 0,
\qquad \D^2 = \DaDa,\qquad \bD^2 = \DbDb = (\D^2)^\dag,\label{lincon}\eeq
where $\Da$ is a spinorial derivative and $R = \bR^\dag$ is the 
auxiliary field of the supergravity multiplet whose \vev\, determines
the gravitino mass: $\langle\l R\r\rangle = \half m_{3\over2}$.
The action written in terms of $L$ is related
to the action written in terms of $S$ by a superfield duality 
transformation; the standard derivation of the duality transformation
requires that $\Omega$ be independent of $S$.  It was shown in
appendix E of the first reference in~\cite{bg} that the the duality
transformation still goes through with a slight modification if this
is not the case.  On the  other hand it might perhaps be reasonable to
impose 
\beq {\pp\Omega\over\pp S} = 0,\label{omegacon}
\eeq
\noindent 
which is in fact the case for the chiral anomaly found in the string
calculation of~\cite{ss}.  We show that it is possible to eliminate
all terms that depend on the full K\"ahler potential $K$, as well as
all terms nonlinear in $F$, and to reproduce the result given
in~\cite{ss}.  However, as discussed in \myapp{dterms}, there may be
a residual S-dependent contribution of the part of the ``D-term''
anomaly that arises from uncancelled logarithmic divergences.

In the following section we outline the four-dimensional Green-Schwarz
mechanism.  In Section 3 we briefly recall the results of~\cite{bg}
and the differences obtained with the present approach. In Sections
4 and 5 we introduce the relevant set of PV fields, outline the
conditions for cancellation of ultraviolet divergences and present
our results for $Z_3$ and $Z_7$ orbifolds.  We summarize our results
in Section 6.  Some details are relegated to Appendices.

\section{The 4-d Green-Schwarz mechanism}
\setcounter{equation}{0}

The four dimensional version of the Green-Schwarz (GS) mechanism was
originally formulated~\cite{prev} as a means of canceling the anomaly
term quadratic in Yang-Mills fields, using the chiral formulation for
the dilaton.  The classical Lagrangian for the Yang-Mills field strength
reads
\beq \L_{\rm YM} = - \sqrt{g}{s\over8}\sum_a\(F^a_{\mu\nu} 
- i\tF^a_{\mu\nu}\)F_a^{\mu\nu}\hc,\qquad s = \l S\r.\label{lym}\eeq
Under the anomalous modular transformation \myref{modtr} the quantum
corrected Lagrangian varies according to 
\beq \Del\L_{\rm YM} = -{\sqrt{g}\over64\pi^2}\sum_{a,i}F^i\[C_a + \sum_b
\(2q^b_i -  1\)C^b_a\]\(F^a_{\mu\nu} - i\tF^a_{\mu\nu}\)
F_a^{\mu\nu}\hc,\label{dellym}\eeq
\noindent
where $C_a$ is the quadratic Casimir in the adjoint representation of
the gauge group factor $\G_a$ and $C_a^b$ is the Casimir for the
representation of the chiral supermultiplet $\Phi^b$.  In $Z_3$ and
$Z_7$ orbifolds one has the universality condition:
\beq C_a + \sum_b\(2q^b_i -  1\)C^b_a = 8\pi^2b\quad\forall\quad i,a,
 \label{univ}\eeq
with $b = 30/8\pi^2$ in the absence of Wilson lines.  The dilaton
is classically invariant under the modular transformation \myref{modtr}.
However if we impose the transformation property:
\beq\Del s = - b F = - b\sum_i F^i(t^i),\qquad
t^i = \l T^i\r,\label{dels}\eeq 
the variation of the classical Lagrangian \myref{lym} cancels
\myref{dellym}.

Now consider the superspace 
Lagrangian\footnote{We use the K\"ahler superspace formulation of
supergravity~\cite{linear}.}
\beq \L =  \superint E\(S + \S\)\Omega = -{1\over8}\superint{E\over R}
\(\bD^2 - 8R\)\(S\Omega\)\hc = - {1\over8}\superint{E\over R}S\Phi\hc,
\label{superl}\eeq
where $E$ is the superdeterminant of the supervielbein, $\Omega$ is the
real superfield appearing in \myref{lincon}, $\Phi$ is its chiral
projection:
\beq \(\bD^2 - 8R\)\Omega = \Phi,\label{defphi}\eeq
and we used superspace integration by parts~\cite{linear}. When
$\Phi$ is replaced by the Yang-Mills superfield strength bilinear 
$\WaWa$, \myref{superl} is just the Yang-Mills Lagrangian that includes
the term in \myref{lym}.  If, under the modular transformation 
\myref{modtr} the quantum Lagrangian varies according to
\beq\Del\L_{\rm anom} = b\superint\[F(T) + \bF(\T)\]\Omega = - {b\over8}
\superint{E\over R}F(T)\Phi\hc,\label{delL}\eeq
the  full Lagrangian is invariant provided 
\beq\Del S = - b F(T).\label{delS}\eeq
However the classical K\"ahler potential for the dilaton is no longer
invariant and must be modified:
\beq k_{\rm class}(S,\S) = -\ln(S + \S) \to k(S,\S) =  -\ln(S + \S
+ V_{G S}), \label{kdil}\eeq
where $V_{G S}$ is a real function of the chiral supermultiplets that
transforms under \myref{modtr} as
\beq \Del V_{G S} = b\(F + \bF\).\label{delV}\eeq
A simple solution consistent with string calculation results~\cite{prev} 
is
\beq V_{G S}  = b g(T,\T),\label{defV}\eeq
where
\beq g(T,\T) = \sum_i g^i(T^i,\T^i),\qquad g^i = - \ln(T^i + \T^i)
\label{kmod} \eeq
is the K\"ahler potential for the moduli.  The modification \myref{kdil}
is the 4d GS term in the chiral formulation.

The 4d GS mechanism is in fact more simply formulated in the linear
multiplet formalism for the dilaton.  The linear superfield $L$
remains invariant, its K\"ahler potential is unchanged, and one simply
adds a term to the Lagrangian.  Using \myref{lincon} and \myref{defphi}:
\bea \L_{G S} &=& - \superint E L V_{G S},\nnn
\Del\L_{G S} &=& - b\superint E L F\hc = {b\over8}\superint{E\over R}
F\(\bD^2 - 8R\)L\hc \eee 
{b\over8}\superint{E\over R}F\Phi\hc = - \Del\L_{\rm anom}\label{lings}
\eea

\section{The anomaly in supergravity}
\setcounter{equation}{0} As mentioned in the introduction, the
quadratic and logarthmic divergences of supergravity can be
cancelled~\cite{pv} by a suitable set of Pauli-Villars (PV)
supermultiplets.  It is straightforward to see~\cite{bg}, by an
examination of the quadratic divergences, that not all of these fields
can have large PV masses that are invariant under nonlinear
transformations on the fields that effect a K\"ahler transformation,
such as the modular transformations \myref{modtr}, as we will
illustrate with an example below.

It was shown in~\cite{bg} that modular noninvariant masses can be
restricted to a subset of PV chiral supermultiplets $\Phi^C$ with
diagonal K\"ahler metric:
\beq K(\Phi^C,\bar\Phi^C) = f^C(Z,\Z)|\Phi^C|^2.\label{kphic}\eeq
\noindent
In particular, those PV fields that have superpotential couplings
to light fields and contribute to the renormalization
of the K\"ahler potential can be chosen to have invariant PV masses.
The fields in \myref{kphic} acquire masses through superpotential
terms:
\beq W(\Phi^C,\Phi'^C) = \mu_C\Phi^C\Phi'^C, \label{wphic}\eeq
\noindent
with $\mu_C$ constant (in the absence of threshold corrections, as
for the cases considered here).  We can define a 
superfield\footnote{The constants $\mu_C$ in \myref{wphic} drop out 
of the variation $\Del\L_{\rm anom}$ of the effective action 
\myref{SFtotanom}, and we ignore them throughout.}
\beq \cM^2_C = \exp(K - f^C - f'^C) = \exp(K - 2\f^C), \qquad \f^C =
\half(f^C + f'^C),\label{M2}\eeq
\noindent
whose lowest component $m_C^2 = \l\cM^2_C\r$ is the $\Phi^C,\Phi'^C$
squared mass. Then the anomalous part of the one-loop corrected
supergravity Lagrangian takes the form~\cite{bg}
\bea \L_{\rm anom} &=& \L_0 + \L_1 + \L_r = \superint E\(L_0 + L_1+
L_r\),\label{SFtotanom}\eea
\bea L_0 &=& {1\over8\pi^2}\[\Tr\eta\ln\cM^2\Omega_0 + K\(\Omega_{G B} 
+ \Omega_D\)\],\qquad L_r = - {1\over192\pi^2}\Tr\eta\int
d\ln\cM\Omega_r,\label{totomega}\eea
\noindent
where $\eta = \pm 1$ is the PV signature,
\beq \Omega_0 = - {1\over24}\Omega_{\rm G B} + \Omega_{\rm Y M} 
- {1\over12}G_{\dot\beta\alpha}G^{\alpha\dot\beta} + {1\over3}R\bR
- {1\over48}\(\D^2R + \bD^2\bR\),\label{Omega0}\eeq
\noindent
\bea \Omega_r &=& - {\partial\over\pp\ln\cM}
\[{1\over4}\(\D^2\ln\cM\Dd\ln\cM\Db\ln\cM\hc\)
- 2G_{\alpha\dot\beta}\Dc\ln\cM\Db\ln\cM
\right.\mmm\qquad\l + \(\ln\cM\lbr{1\over8}\bD^2\D^2\ln\cM 
+ \Dc(R\Da\ln\cM)\rbr
\hc\)\right.\mmm\qquad\l + \half\Dc\ln\cM\Da\ln\cM\Dd\ln\cM\Db\ln\cM
\right.\mmm\qquad\l - (\ln\cM)^2\({1\over4}\Dc\La + \ln\cM\Dc\Xa\)\],
\label{Omegar}\eea 
\noindent
with 
\beq \Xa = - {1\over8}\chiproj\Da K,\qquad 
\La = \chiproj\Da\ln\cM, \label{xlalpha}\eeq
\noindent
$G_{\dot\beta\alpha}$ is an auxiliary superfield of the gravity
supermultiplet, and $\Omega_D$ represents the ``D-term'' anomaly (see
\myapp{dterms}) that, together with a contribution to the Gauss-Bonnet
term $\Omega_{\rm G B}$:
\beq \Omega_{\rm G B} = - 8\Omega_W - {4\over3}\Omega_X  
-G_{\dot\beta\alpha}G^{\alpha\dot\beta} + 4R\bR,\label{Omegagb}\eeq
\noindent
arises from uncanceled total derivatives with logarithmically
divergent coefficients as discussed in the introduction. Supersymmetry
of these terms requires a field-dependent cut-off:
\beq \Lambda = \mu_0 e^{K/4}.\label{lambda} \eeq
\noindent
The constant $\mu_0$ drops out of the effective action \myref{SFtotanom}.

The Chern-Simons superfields $\Omega_W$, $\Omega_X$ and 
$\Omega_{\rm Y M}$ are defined by their chiral projections:
\beq \chiproj\Omega_W = W^{\alpha\beta\gamma}W_{\alpha\beta\gamma},
\qquad \chiproj\Omega_X = \Xc\Xa,\qquad\chiproj\Omega_{\rm Y M}
= \WaWa.\label{omegas}\eeq
\noindent
where $W_{\alpha\beta\gamma}$ is the superfield strength for
space-time curvature.

$\L_1$ is defined by its variation:
\beq \Del L_1 = {1\over8\pi^2}{1\over192}\Tr\eta\Del\ln\cM^2\Omega'_L 
= {1\over8\pi^2}{1\over192}\Tr\eta H\Omega'_L\hc,\label{dell1}\eeq
\noindent
where under \myref{modtr} $\ln\cM^2$ transforms as
\beq \Del\ln\cM^2 = H + \bar H,\eeq
with $H$ holomorphic. Defining 
\bea \chiproj\Omega_f &=& \fc\fa,\qquad \chiproj\Omega_{\f} = \bar\fc
\bar\fa,\qquad \chiproj\Omega_{\f X} = \bar\fc\Xa,\nnn
\fa &=& - {1\over8}\chiproj\Da f,\qquad
\bar\fa = - {1\over8}\chiproj\Da\f,\label{Omegaf}\eea
we have
\bea \Omega'_L &=& 192\Omega_f - 128\Omega_{\f} - 64\Omega_{\f X},\nnn
\Del L_1 &=& {1\over8\pi^2}\Tr\eta H\(\Omega_f - {2\over3}\Omega_{\f} -
{1\over3}\Omega_{\f X}\)\hc\label{Omegalprime}\eea
\noindent

The general form of $f^C$ is taken to be 
\bea \ln f^C &=& \alpha^C K(Z,\Z) + \beta^C g(T,\T) + \delta^C k(S,\S)
+ \sum_n q^C_n g^n(T^n,\T^n),\nnn 
\ln\f^C &=& \bar\alpha^C K + \bar\beta^C g
+ \bar\delta^C k + \sum_n\q^C_n g^n,\nnn 
H^C &=& \(1 - 2\bar\gamma^C\)F(T) - 2\sum\q^C_n F^n(T^n),
\qquad \bar\gamma^C = \bar\alpha^C + \bar\beta^C.\label{deffh}\eea
\noindent

The traces in $\Del\L_{\rm anom}$ can be evaluated using only PV fields
with noninvariant masses or using the full set of PV fields, since
those with invariant masses, $H^C=0$, drop out.  The contribution
$\Del L_0$ to the anomaly is linear in the parameters $\alpha^C,\beta^C,
q^C_n$; as a consequence the traces are completely determined by the
sum rules~\cite{pv}
\bea N' &=& \sum_C\eta^C = - N - 29,\qquad N'_G = 
\sum_\gamma\eta_\gamma^V = - 12 - N_G,\nnn
\sum_C\eta^C\ln f^C &=& - 10K - \sum_q q^p_n g^n,\label{sums}\eea
\noindent
that are required to assure the cancellation of all quadratic and
logarithmic divergences.  In \myref{sums} the index $C$ denotes any
chiral PV field, the index $\gamma$ runs over the Abelian gauge PV
superfields that are needed to cancel some gravitational and
dilaton-gauge couplings, and the sum over $p$ includes all the light
chiral multiplet modular weights with $q^S_n = 0,\; q^{T^i}_n =
2\del^i_n$.  $N$ and $N_G$ are the total number of chiral and gauge
supermultiplets, respectively, in the light sector.  All PV fields
with noninvariant masses have $\del = 0$, and most\footnote{There is a
  set of chiral multiplets in the adjoint representation of the gauge
  group that has $\ln f = K - k$; these get modular invariant masses
  though their coupling in the superpotential to a second set with
  $\ln f = k$. These cancel renormalizable gauge interactions and
  gauge-gravity interactions, respectively.  Together with a third
  set, that has $f = 1$ and contributes to the anomaly, they cancel
  the Yang-Mills contribution to the beta-function.\label{foot3}} with
$\del\ne0$ have $\alpha = \beta = q_n = 0$.  For the purposes of the
present analysis we can ignore the latter.

To see that not all the PV chiral multiplets can have invariant
masses, there is a quadratically divergent contribution from the light
sector given by
\beq \L_Q \ni -\l{\sqrt{g}}{\Lambda^2\over64\pi^2}\(3 + N_G - N\)\Dc\Xa
\r,\label{quad1}\eeq
\noindent 
where $\Xa$ is defined in \myref{xlalpha}.  The Pauli-Villars
contribution to the operator in \myref{quad1} is
\beq \L^{\rm P V}_Q \ni -\l{\sqrt{g}}{\Lambda^2\over64\pi^2}\(N'_G - N'
- 2\alpha\)\Dc\Xa\r, \label{quad2}\eeq
\noindent 
where $\alpha = \sum\eta^C \alpha^C$. The PV chiral multiplets include
a subset $\theta^a$ with $N'_\theta = N'_G$ which form massive vector
supermultiplets with the PV Abelian gauge supermultiplets; these
cancel in \myref{quad2}.  The remainder get superpotential masses as
in \myref{wphic}. The pair $\Phi^C,\Phi'^C$ will have an invariant
mass if $\ln\f^C = \bar\alpha^C = \half$, in which case the total
contribution of the pair to \myref{quad2} vanishes
identically. Therefore chiral fields with noninvariant masses are
needed to cancel \myref{quad1}.

In contrast to $\L_0$, the contributions to the anomaly from $\L_1$
and $\L_r$ are nonlinear in the parameters $\alpha,\beta,q$, and
depend on the details of the PV sector.  In~\cite{bg} the PV sector
was constructed in such a way that
\beq f^C = f'^C = \f^C\label{samef}\eeq
\noindent 
for the PV fields with noninvariant masses.  In this case
\myref{Omegalprime} reduces to
\beq (\Omega'_L)_{[1]} = 64\(\Omega_{\f} - \Omega_{\f X}\) =
\Omega_L - 16\Omega_X,\qquad \chiproj\Omega_L = \Lc\La,
\label{OprimeL}\eeq
\noindent
and, for example,
\bea \Tr\eta H\Omega_{\f} &=& \sum_C\eta_C\[\(1 - 2\bar\gamma^C\)F -
2\sum_n\q^C_n F^n\]\times\ddd
\(\bar\alpha^C\Xc + \bar\beta^C\gc + \sum_m\q^C_m
\gc_m\)\(\bar\alpha^C\Xa + \bar\beta^C\ga + \sum_l\q^C_l
\ga^l\).\label{trHOf}\eea
The Pauli-Villars modular weights $q^C_n$ are related to the modular
weights $q^p_n$ of the light fields by the conditions for the 
cancellation of UV divergences. In the $Z_3$ and $Z_7$ orbifolds
considered below, the latter satisfy sum rules of the form:
\beq \sum_p q^p_n = A_1, \qquad \sum_p q^p_n q^p_m = A_2 + B_2\del_{m n}.
\label{qsums}\eeq
\noindent
The first sum rule in \myref{qsums} assures the university of the anomaly
proportional to $\Omega_0 - \Omega_{\rm Y M}$.  However, in the PV
sector used in~\cite{bg} the second equality led to a nonuniversal
term:
\bea \Tr\eta H\Omega_{\f} &\ni& - 4\sum_{p,m,n}q^p_n q^p_m F^n
\gc_m\(\bar\alpha^C\Xa + \bar\beta^C\ga\) \eee
- 4\(A_2 F\gc + B_2 F^n\gc_n\)\(\bar\alpha^C\Xa + \bar\beta^C\ga\).
\label{bad}\eea
The sum rule cubic in the modular weights is more complicated, but in
general leads to additional nonuniversal terms.  These can be avoided
by imposing $\q^C_n = 0$ for fields with noninvariant masses, but if
\myref{samef} is imposed we get 
\beq\Tr\eta H\Omega'_L = F\(a\Xc\Xa + b\Xc\ga + c\gc\ga\),\eeq
\noindent
which does not include the term proportional to
\beq F\sum_n\gc_n\ga^n\label{SSanom} \eeq
\noindent
found in the string calculation\footnote{In fact the
  four-form $\epsilon^{\mu\nu\rho\sigma}g^n_{\mu\nu}g^n_{\rho\sigma}$
  with $g^n_{\mu\nu} = (\pp_\mu t^n\pp_\nu\t^{\n})g^n_{t^n\t^{\n}} -
  (\mu\leftrightarrow\nu)$, that appears in the chiral part of
  \myref{SSanom}, vanishes identically.  We find it curious that the
  authors of~\cite{ss} neglected to comment on this fact.  However
the associated conformal anomaly is nontrivial.} 
of~\cite{ss}.\label{foot1}

In the following we relax the assumption \myref{samef}, impose
$\q^C_n = 0$, but with $q^C = - q'^C \ne 0$.  This still assures a
universal anomaly, but allows more freedom in determining its
coefficient; in particular, we are able to reproduce the term
\myref{SSanom}.

\section{Cancellation of UV divergences}
\setcounter{equation}{0} The full set of PV fields needed to regulate
light field couplings is described in Section 3 of~\cite{bg}.  Among
those, here we are primarily concerned with the set $\dZ^P =
\dZ^I,\dZ^A$, with negative signature, $\eta^{\dZ} = -1,$ that
regulates most of the couplings, including all renormalizable
couplings, of the light chiral supermultiplets $Z^p = T^i,\Phi^a$.
Covariance of the $\dZ^P$ K\"ahler metric requires that these fields
transform under \myref{modtr} like $d Z^p$:
\beq \dZ'^I = e^{-2F^i}\dZ^I,\qquad \dZ'^A = e^{- F^a}\(\dZ^A - \sum_j 
F^a_j\Phi^a\dZ^J\),\qquad F^a = \sum_i F^i(T^i)\label{dZtr}\eeq
\noindent
Invariance of the full PV K\"ahler potential for the
$\dZ^P$ and covariance of their superpotential under \myref{modtr}:
\beq K(\dZ') = K(\dZ), \qquad W(\dZ') = e^{-F(T)}W(\dZ),\eeq
\noindent 
can be made manifest if we supplement~\cite{bg} these fields
with three additional PV fields $\dZ^N,\;N = 1,2,3$,
with K\"ahler potential
\beq K(\dZ^N) = \sum_{i=n}\left|\dZ^N + \dot a \chi^n(T^i)\dZ^i\r^2,
\qquad \chi^n(T'^i) = e^{2F^n}\(\chi^n(T^i) + F^n_i\),
\label{KdZ}\eeq
\noindent
and that transform under \myref{modtr} according to
\beq \dZ'^N = \dZ^N - \dot a F^n_i(T^i)\dZ^I,\label{dZNtr}\eeq
\noindent
where $\dot a$ is a nonzero constant.\footnote{Depending on the choice
  of the functions $\chi^n(T^i)$, one might need to
  introduce~\cite{bg} several copies of the sets $\dZ_\lambda^{P,N}$,
  with constraints on the parameters $\dot a_\lambda$ in such a way
that no new divergences are introduced by the fields 
$\dZ^N$.}\label{foot2}

We wish to give these PV fields modular invariant
masses.  The simplest way to do this is to introduce fields
$\dY_P,\dY_N$ with the same signature, opposite gauge charges and the
inverse K\"ahler metric.  However this would have the effect of
canceling the $\dZ$ contributions that are linear in the generalized
field strength 
\beq G_{\mu\nu} = [D_\mu,D_\nu],\label{gmn}\eeq
and doubling the quadratic $\dZ$ contributions.  Instead we introduce
fields $\dY_P,\dY_N$ with with gauge charges
\beq (T_a)_{\dY} = - (T_a^T)_{\dZ} = - (T_a^T)_Z,\label{Ta}\eeq
\noindent
and K\"ahler potential 
\bea K(\dY) &=& e^{\dot G}\(\sum_A e^{- g^a}|\dY_A|^2 
+ \sum_I e^{- 2g^i}|\dY_I - \dot a\chi^n(T^i)\dY_N|^2 + |\dY_N|^2\),\nnn 
g^a &=& \sum_n q_n^a g^n, \qquad \dot G = \dot\alpha K + \dot\beta g,
\qquad \dot\alpha +\dot\beta = 1.\label{KdY}\eea
\noindent
\myref{KdY} is modular invariant, and the PV mass superpotential 
\beq W(\dZ,\dY) = \dot\mu\(\dZ^A - \dot a^{-1}q^a_n\Phi^a\dZ^N\)\dY_A
+ \sum_{i = n}\dot\mu_n\(\dZ^I\dY_I + \dZ^N\dY_N\),\label{WZY}\eeq
is covariant, provided under \myref{modtr}
\beq \dY'_A = e^{-F + F^a}\dY_A,\qquad \dY'_I = e^{- F + 2F^i}
\(\dY_I + \dot a F^n_i\dY_N\),
\qquad \dY'_N = e^{- F}\dY_N.\label{dYtr}\eeq

It remains to cancel the divergences introduced by the fields $\dY$.
This was achieved in~\cite{bg} by an additional set of chiral PV
fields, collectively called $\Psi$, with diagonal metric \myref{kphic},
superpotential  \myref{wphic}, with prefactors \myref{deffh} satisfying
\myref{samef} and $\alpha^\Psi = \delta^\Psi = 0$.  In addition
$\dot\alpha=0$ in \myref{KdY} was assumed.  Here we use a different
set of fields, for which we assume only $\del^C$ = 0, as well as
allowing $\dot\alpha\ne 0$.  For this reason we also include in
the present analysis the set of fields $\phi^C$ with prefactors 
\beq \ln f^{\phi^C} = {\alpha^C K} \label{fphic}\eeq
\noindent
that regulate certain gravity supermultiplet loops.  These must be
included together with the PV fields introduced below in implementing
the sum rules \myref{sums}.  We take the following set:
\bea \Phi^P: && \ln f^{\Phi^P} = \sum_n q^P_n g^n = \alpha^\Phi K +
\beta^\Phi g - \ln f^{\Phi'^P},\qquad \alpha^\Phi + \beta^\Phi =
1,\nnn \psi^{P n}: && \ln f^{P n} = \alpha_\psi^P K + \beta_\psi^P g +
q_\psi^P g^n, \qquad \alpha_\psi^P + \beta_\psi^P = \gamma_\psi^P,
\qquad \bar q_\psi^P = 0,\nnn T^P: && \ln f_T^P = \alpha_T^P K +
\beta_T^P g, \qquad \alpha_T^P + \beta_T^P =
\gamma_T^P.\label{pvfields}\eea 
\noindent 
The pairs $\Phi^P,\Phi'^P$ have modular invariant masses and do not
contribute to the anomaly, but they play an important role in
canceling certain divergences. In the case of $Z_7$ orbifolds we take
them to be charged under the two \uo's of that theory.  They have no
other gauge charges, the $\psi^{P n}$ are taken to be gauge neutral,
and the $T^P$ have a priori arbitrary gauge charges.  For those in
real representations of the gauge group one can take $T^P = T'^P$.
In \myapp{uv} we display a simple solution to the constraints with
some $T$'s in the fundamental and antifundamental representation of 
the non-Abelian gauge group factors, some with $U(1)$ charges in the
$Z_7$ case, and some gauge singlets.

The quadratic and logarithmic divergences we are concerned with here
involve the superfield strengths $-i(T_a)\Wa^a$, $X_\alpha$ and
\beq \Gamma^C_{D\alpha} = - {1\over8}\chiproj\Da Z^i\Gamma^C_{D i},
\label{gamma}\eeq
\noindent
associated with the Yang-Mills, K\"ahler and reparameterization
connections, respectively.  Since the theories considered here have no
gauge anomalies, cancellation of quadratic divergences requires
\beq \Tr\eta\Gamma_\alpha = 0,\label{quad}\eeq
\noindent
and cancellation of logarithmic divergences requires 
\beq \Tr\eta\Gamma_\alpha\Gamma_\beta = \Tr\eta\Gamma_\alpha T^a =
\Tr\eta(T^a)^2 = 0,\label{log}\eeq
\noindent
where $\eta = + 1$ for light fields.  Cancellation of all
contributions linear and quadratic in $\Xa$ is assured by the
conditions in \myref{sums} together with \myref{a5} of \myapp{uv}.
The Yang-Mills contribution to the term quadratic in $\Wa$ is canceled
by chiral fields in the adjoint (see footnote on page~\pageref{foot3})
that we need not consider here.  Finally, cancellation of linear
divergences requires cancellation of the imaginary part of
\beq \Tr\eta X_\chi = \Tr\eta\phi G\cdot\tG, \qquad \tG^{\mu\nu}= \half
\epsilon^{\mu\nu\rho\sigma} G_{\rho\sigma},\label{defX}\eeq
\noindent
where $G_{\mu\nu}$ is the field strength associated with the fermion
connection;\footnote{Here we neglect the spin connection which is 
considered in \myapp{dterms}.} for left-handed fermions:
\beq G_{\mu\nu} = - \Gamma^C_{D\mu\nu} + iF^a_{\mu\nu}(T_a)^C_D
+ \half X_{\mu\nu}\del^C_D,\label{defG}\eeq
\noindent
and for a generic PV superfield $\Phi^C$ with diagonal metric, its
fermion component $\chi^C$ transforms under \myref{modtr} as
\beq \chi'^C = e^{\phi^C}\chi^C,\qquad \phi^C = \(\half - \alpha^C - 
\beta^C\)F - \sum_i F^i(T^i)q^C_i.\label{defphic}\eeq

The full set of conditions is extensive, and we evaluate them in
\myapp{uv}.  In this section we simply outline how to
obtain a universal anomaly using PV regularization.  For this purpose
we focus on terms contributing to UV divergences that could 
potentially spoil universality.  An important feature in our results
is the fact that the expression
\beq \epsilon^{\mu\nu\rho\sigma} g^i_{\mu\nu}
g^{i}_{\rho\sigma} = 0,\label{zero}\eeq
\noindent
vanishes identically, and the expressions
\bea X^{i j} &=& \epsilon^{\mu\nu\rho\sigma}\im F^i g^i_{\mu\nu}
g^{j\ne i}_{\rho\sigma} =
4\epsilon^{\mu\nu\rho\sigma}\im F^i\pp_\mu g^i_\nu
\pp_\rho g^j_\sigma = 4\pp_\rho\(\epsilon^{\mu\nu\rho\sigma}\im F^i
\pp_\mu g^i_\nu g^j_\sigma\),\nnn
X^i &=& \half\epsilon^{\mu\nu\rho\sigma}\im F^i g^i_{\mu\nu}
X_{\rho\sigma} =
4i\pp_\rho\(\epsilon^{\mu\nu\rho\sigma}\im F^i
\pp_\mu g^i_\nu\Gamma_\sigma\),\nnn
X^{i a} &=& \epsilon^{\mu\nu\rho\sigma}\im F^i g^i_{\mu\nu}
F^a_{\rho\sigma}
= 4\pp_\rho\(\epsilon^{\mu\nu\rho\sigma}\im F^i
\pp_\mu g^i_\nu A^a_\sigma\),\label{epsilons}\eea  
\noindent
are total derivatives, where $A_\mu^a$ is an Abelian gauge field,   
\beq g^i_\mu = -{\pp_\mu t^i - \pp_\mu \t^{\ibar}\over t^i
+\t^{\ibar}},\qquad g^i_{\mu\nu} = \pp_\mu g^i_\nu - \pp_\nu g^i_\mu,
\qquad \Gamma_\mu = {i\over4}\(\D_\mu z^i K_i - \D_\mu\z^{\m}K_{\m}\),
\label{gdefs}\eeq
\noindent
and $X_{\mu\nu} = 2i\(\pp_\mu\Gamma_\nu - \pp_\nu\Gamma_\mu\)$ is
defined in \myref{xmunu}.  

If, for example, we replaced $g^n(T^n,\T^{\n})$ everywhere by the
K\"ahler potential for the $n$th untwisted sector, a possibility
considered in~\cite{bg}, the above would not hold, and we would be
unable to obtain a universal anomaly coefficient.  Specifically, we
would be not be able to cancel the terms cubic in $q^p_n$ that appear
in $X_\chi^{\dY}$, suggesting that the present construction is the
only viable possibility.  This agrees with the results of~\cite{ss},
where it was found that the untwisted K\"ahler moduli are the only
chiral supermultiplets that appear in the chiral anomaly (see however
footnote page~\pageref{foot1}). 

\subsection{Reparameterization curvature terms}
The functions $\chi^n(T^i)$ in \myref{KdZ} and \myref{KdY} do not
contribute to the quantities in \myref{gamma} and \myref{defG} (see
footnote page~\pageref{foot2}), and, using \myref{qsums}, one obtains
\bea \Tr\deta\Gamma_\alpha^{\dY} &=& - \[(N + 2)\dot\beta - A_1\]\ga,
\nnn \Tr\deta\Gamma^{\dY}_\alpha\Gamma^{\dY}_\beta &=&
- 2\dot\alpha\[\dot\beta(N + 2) - A_1\]X_\alpha g_\beta - 
\[\dot\beta^2(N + 2) - \dot\beta A_1 + A_2\]
g_\beta g_\alpha\ddd - B_2\sum_n g^n_\alpha g^n_\beta.\label{log3}\eea
\noindent
In addition we have
\bea X^{\dY}_\chi &=& \half(N + 2)F\dot G\cdot\tilde{\dot G} -
F\sum_{p,n}q^p_n\dot G\cdot\tg^n + 
\half F\sum_{p,m,n}q^p_m q^p_n g^m\cdot\tg^n\ddd 
- \sum_{p,n}q^p_n F^n\dot G\cdot\tilde{\dot G} + 2\sum_{p,n,m}
q^p_m q^p_n F^m\dot G\cdot\tg^n - \sum_{p,l,m,n}q^a_l q^a_m q^a_n F^l 
g^m\cdot\tg^n,\label{cubic}\eea
\noindent
In addition to the sum rules listed in \myref{qsums}, we have
\bea \sum_p q^p_i q^p_j q^p_k &=& A_3 + B_3\del_{i j}\del_{i k}
+ C_3\[\del_{i j}\(\del^1_i\del^2_k + \del^2_i\del^3_k
+ \del^3_i\del^1_k\) + {\rm cyclic}(i j k)\]\ddd
+ D_3\[\del_{i j}\(\del^2_i\del^1_k + \del^3_i\del^2_k + 
\del^1_i\del^3_k\) + {\rm cyclic}(i j k)\],\label{3q}\eea
\noindent
with $C_3 = D_3$ for $Z_3$, $C_3 \ne D_3$ for $Z_7$. 
Then using \myref{zero} and \myref{epsilons}, \myref{cubic} reduces to
\bea X^{\dY}_\chi &=& \half(N + 2)F\dot G\cdot\widetilde{\dot G} 
- A_1F \dot G\cdot\tg + 
\half A_2F g\cdot\tg - A_1F\dot G\cdot\widetilde{\dot G}\ddd 
+ 2A_2F\dot G\cdot\tg - A_3F g\cdot\tg  + \mbox{total derivative}.
\label{cubic2}\eea
\noindent
Since $\q^\Phi = \q^\psi = 0$, terms cubic in these modular weights do not
contribute to $X_\chi^\Phi,X_\chi^\psi$.  Further, since
\beq q^{P n}_m q^{P n}_l = (q^P_\psi)^2\del^n_m\del^n_l,\label{q2psi}\eeq
\noindent
there are no contributions to $X_\chi^\psi$ quadratic in $\psi$
modular weights, and since $q^P_\psi$ is independent of $n$,
$X_\chi^\psi$ depends only on $F,g_{\mu\nu}$ and $X_{\mu\nu}$.  Then
imposing
\beq \sum_P\eta^P q^P_n = a_1,\qquad
\sum_P\eta^P q^P_n q^P_m = a_2 + b_2\del_{m n},
\label{Phiqsums}\eeq
\noindent
$X_\chi^\Phi$ can also be made to depend only on $F,g_{\mu\nu}$
and $X_{\mu\nu}$.  The terms linear in the $\Phi$ and $\psi$ modular
weights drop out of $(\Tr\eta\Gamma_\alpha)_{\Phi,\psi}$, and one 
obtains\footnote{In \myref{Phiqsums} and for $\Phi$ in \myref{gamma2Pp},
the sum is over $P$ only, while for $\psi$, $\sum_P \equiv\sum_P +
\sum_{P'}$, since $P$ and $P'$ are interchangeable in the latter,
but not in the former.}
\bea \(\Tr\eta\Gamma_{\alpha}\Gamma_{\beta}\)_\Phi &=&
\sum_P\eta_\Phi^P\Ga^\Phi G_\beta^\Phi - a_1\(\Ga^\Phi g_\beta +
\ga G^\Phi_\beta\) + 2a_2\ga g_\beta + 2b_2\sum_n\ga^n g_\beta^n,\nnn
 G^\Phi &=& \alpha^\Phi K + \beta^\Phi g,\nnn
\(\Tr\eta\Gamma_{\alpha}\Gamma_{\beta}\)_\psi &=&
\sum_P\eta_\psi^P\[3\Ga^P G_\beta^P + q_\psi^P\(\Ga^P g_\beta +
\ga G^P_\beta\)\] + B_\psi\sum_n\ga^n g^n_\beta,\nnn
G^P &=& \alpha ^P K + \beta^P g,\qquad B_\psi =
\sum_P\eta_\psi^P(q^P_\psi)^2.\label{gamma2Pp}\eea
\noindent
To cancel the last term in \myref{log3} we require
\beq 2b_2 + B_\psi = B_2.\label{gngn}\eeq
\noindent
The $\dY$ and $\Phi$ fields do not contribute to the anomaly, and
the coefficient of the term \myref{SSanom} is determined by $B_\psi$.
The remaining terms in \myref{log3} and \myref{cubic}
can be cancelled by a combination of the full set of PV fields
in \myref{fphic} and \myref{pvfields}, as shown in \myapp{uv}.

\def\ua{$U(1)_a$}

\subsection{Yang-Mills field strengths}
The gauge charges\footnote{We use the standard charge normalization
  such that \myref{univ} is satisfied with $C^b_a = (\Tr
  T_a^2)_{R(b)}$, where $R(b)$ is the gauge group representation of
  the chiral supermultiplet $\Phi^b$; this differs by a factor
  $\sqrt{2}$ from the normalization used in~\cite{ss}.}  and modular
weights in $Z_3$ and $Z_7$ orbifold compactifications without Wilson
lines are given in~\cite{ss} and Appendix D.5 of~\cite{bg}.  The
universality of the anomaly term quadratic in Yang-Mills fields
strengths is guaranteed by the universality condition \myref{univ}, as
illustrated in \myapp{uv}.  Since gauge transformations commute with
modular transformations, a set of chiral multiplets $\Phi^b$ that
transform according to a nontrivial irreducible representation $R$ of
a non-Abelian gauge group factor $\G_a$ have the same modular weights
$q^R_n$ such that 
\beq \sum_{b\in R}q^b_n(T_a)^b_b = q^R_n(\Tr T_a)_R = 0.\eeq  
\noindent
Therefore terms linear in Yang-Mills field strengths occur only for
Abelian gauge group factors.  There are none in $Z_3$, but two in $Z_7$,
which we refer to as \ua, $a = 1,2$, with charges $Q_a$.  These are
anomaly free; their traces vanish when taken over the full spectrum of
chiral multiplets.  Defining
\beq Q_{a n} = \sum_b q_n^b Q_a^b, \qquad Q_{a n m} = 
\sum_b q_n^b q_m^b Q_a^b,\label{defQ}\eeq
\noindent
we have for $Z_7$:
\bea Q_{1 n} &=& \half(8,2,-10),\qquad Q_{2n} = 
{1\over2\sqrt{3}}(12,-18,6), \qquad n = (1,2,3),
\nnn Q_{1n m} &=& \half(5,-4,-1),\qquad Q_{2n m} = 
{1\over2\sqrt{3}}(-3,-6,9), \mmm\qquad\quad n m = (12,23,31).\label{Qs}
\eea
\noindent
These satisfy
\beq \sum_n Q_{a n} = 0,\qquad Q_{a n m} = - \half|\epsilon_{n m l}
|Q_{a l}.\label{Qconds}\eeq
\noindent
We wish to cancel the $\dY$-loop contribution to logarithmic
divergences 
\beq  \(\Tr\eta\ga^n T_a\)_{\dY} = - \sum_b q_n^b Q_a^b g^n_\alpha = -
Q_{a n}g^n_\alpha,\label{xan}\eeq
and, dropping terms proportional to the last expression in \myref{epsilons},
$\dY$ contributions to linear divergences:
\bea X^{\dY}_\chi &\ni& 
\sum_b\tF^a_{\mu\nu}Q^b_a\[g^{n\mu\nu}q_n^b\(F - 2q^b_m F^m\)
+ 2q^b_n F^n\lbr\(\dot\alpha - \half\)X^{\mu\nu} + \dot\beta g^{\mu\nu}
\rbr\]\eee
\tF^a_{\mu\nu}F^1\lbr\(Q_{a2}g^{2\mu\nu} + Q_{a3}g^{3\mu\nu}\) + 2
Q_{a1}\[\(\dot\alpha - \half\)X^{\mu\nu} + \dot\beta\(g^{2\mu\nu} 
+ g^{3\mu\nu}\)\]\right.\mmm \l 
- 2\(Q_{a12}g^{2\mu\nu} + Q_{a13}g^{3\mu\nu}\)\rbr
 + \mbox{cyclic (1,2,3) + total derivative}.\label{xmn}\eea
\noindent
Using \myref{Qconds}, \myref{xmn} becomes
\bea X_\chi^{\dY} &\ni& \tF^a_{\mu\nu}F^1\lbr\(Q_{a2}g^{2\mu\nu} +
 Q_{a3}g^{3\mu\nu}\)  + 2Q_{a1}\[\(\dot\alpha - \half\)X^{\mu\nu} 
+ \dot\beta\(g^{2\mu\nu} + g^{3\mu\nu}\)\]\right.\mmm\l 
+ \(Q_{a3}g^{2\mu\nu} + Q_{a2}g^{3\mu\nu}\)\rbr
+ \mbox{cyclic (1,2,3) + total derivative}\eee
\tF^a_{\mu\nu}F^1\[\(Q_{a2} + Q_{a3} + 2\dot\beta Q_{a1}\)
\(g^{2\mu\nu} + g^{3\mu\nu}\) + Q_{a1}\(2\dot\alpha - 1\)X^{\mu\nu}\]
\mmm + \mbox{cyclic (1,2,3) + total derivative}.\eea

Now we assign \ua\, charges $Q^P_a$ and $-Q^P_a$ to $\Phi^P$ and
$\Phi'^P$, respectively.  This gives a contribution to logarithmic
divergences
\beq 2\sum_P\eta^P q_n^P Q_a^P g^n_\alpha \equiv
Q^\Phi_{a n}g^n_\alpha.\label{xan2}\eeq
Cancellation of \myref{xan} requires
\beq Q^\Phi_{a n}= Q_{a n},\eeq
The $\Phi$ contribution to linear divergences is 
\bea X_\chi^\Phi &\ni& -2\sum_P\eta^P Q^P_a q^P_n\tF^a_{\mu\nu}
F^n\[(\alpha^\Phi - 1)X^{\mu\nu} + \beta^\Phi g^{\mu\nu}\]\eee
- Q^\Phi_{a1}\tF^a_{\mu\nu}F^1
\[(\alpha^\Phi - 1)X^{\mu\nu} + \beta^\Phi\(g^{2\mu\nu} + g^{3\mu\nu}\)\]
\mmm + \mbox{cyclic (1,2,3) + total derivative}.\eea
To cancel the $X^{\mu\nu}$ term we require  
\beq \dot\alpha = \half\alpha^\Phi, \qquad \dot\beta = 
1 -\half\alpha^\Phi = \half\beta^\Phi + \half.\eeq
Then
\bea X_\chi^{\dY} &\ni& 
\tF^a_{\mu\nu}F^1\lbr\[Q_{a2} + Q_{a3} + \(\beta^\Phi + 1\)Q_{a1}\]
\(g^{2\mu\nu} + g^{3\mu\nu}\) + Q_{a1}\(\alpha^\Phi - 1\)X^{\mu\nu}\rbr
\mmm + \mbox{cyclic (1,2,3) + total derivative}\eee
\tF^a_{\mu\nu}F^1Q_{a1}\[\beta^\Phi 
\(g^{2\mu\nu} + g^{3\mu\nu}\) + \(\alpha^\Phi - 1\)X^{\mu\nu}\]
\mmm + \mbox{cyclic (1,2,3) + total derivative} = - X_\chi^\Phi,\eea
\noindent
up to a total derivative.

Note that this is a highly nontrivial result.  In addition to the
importance of the properties in \myref{epsilons}, the relations
\myref{Qconds}, that are specific to the $Z_7$ orbifold we are
considering, are crucial to the cancellations in this section. Since
the $\Phi$ have modular invariant masses, the $\psi$'s have no gauge
charges, and the $T$'s have $n$-independent prefactors $f^T$, no terms
linear in the gauge field strengths appear in the anomaly.

Finally we remark that a pair of PV fields $\Phi^C,\Phi'^C$ with
superpotential coupling \myref{wphic} contributes an amount
\beq \(\phi^C + \phi'^C\)C^C_a = \Del\cM^2C^C_a\label{XFF}\eeq
\noindent 
to the coefficient of $F^a\cdot\tF_a$ in \myref{defX}. This vanishes 
for pairs with invariant masses, and its form assures that the anomaly arising
from PV masses in the regulated theory matches the anomaly due to linear
divergences in the unregulated theory.  In particular it makes no 
difference whether or not we assign non-Abelian gauge charges to the
$\Phi^P$, and their \ua\, charges have no affect on the term in the 
anomaly quadratic in the \ua\, field strengths.

\section{The anomaly in $Z_3$ and $Z_7$ orbifolds}
\setcounter{equation}{0}
In \myapp{uv} we show that is possible to cancel all the ultraviolet
divergences from the $\dY$ fields with a simple choice of the set 
\myref{pvfields} such that the fields with noninvariant masses have
the properties
\beq\Tr\eta(\ln\cM)^{n>1} = \Del\Tr\eta(\ln\cM)^{n>1} = 
\Tr\eta(\Del\ln\cM)(\f_\alpha)^{n>0} =0,\label{zeros}\eeq
\noindent
and the anomaly due to the variation of \myref{SFtotanom} reduces to
\bea \del\L_{\rm anom} &=& b\int d^4\theta E F\Omega,\nnn
\Omega &=& \Omega_{\rm Y M} - {1\over24}\Omega_{\rm G B} 
- {b_{\rm spin}\over48b}\(4G_{\dot\beta\alpha}G^{\alpha\dot\beta} 
- 16R\bR + \D^2R + \bD^2\bR\) + {1\over30}\(\Omega_f + \Omega_D\),
\label{finalan}\eea
\noindent
where (see \myapp{dterms})
\beq 8\pi^2b_{\rm spin} = 8\pi^2b + 1 = 31.\label{bspin}\eeq
\noindent
The results for the Gauss-Bonnet and Yang-Mills terms are 
well-established~\cite{prev} and result from the universality conditions
\myref{univ} and \myref{univ2}, as illustrated in the appendices. The
only other term in \myref{finalan} that contains a chiral anomaly is
$\Omega_f$, which, using the set \myref{pvfields} of PV fields,
is a priori a product of the chiral superfields $\Xa,\;\ga$
and $\ga^n$.  We show in \myapp{uv} we may choose the PV parameters
such that 
\beq \Omega_f = 30\sum_n\gc_n\ga^n,\label{omegaf}\eeq
\noindent
in agreement with the string callculation of~\cite{ss}.

The anomaly is canceled provided the Lagrangian for the dilaton $S,\S$
is specified by the coupling \myref{superl} and the K\"ahler potential
\myref{kdil}, or, equivalently, the linear supfield $L$ satisfies 
\myref{lincon} and the GS term \myref{lings} is added to the Lagrangian.

\section{Conclusions}
\setcounter{equation}{0}

We have shown that a suitable choice of Pauli-Villars regulator fields
allows for a full cancellation of the chiral and conformal anomalies
associated, respecively, with the linear and logarithmic divergences
in the effective supergravity theories from $Z_3$ and $Z_7$
compactification of the weakly coupled heterotic a string without
Wilson lines.  In particlar we were able to reproduce the form of
the chiral anomaly found in a string theory calculation~\cite{ss} for
these two models.

In a future study we will extend our analysis to an example of $Z_3$
orbifold compactification with Wilson lines and an anomalous \uo.

\vskip .3in
\noindent{\bf Acknowledgments.}  We are happy to acknowledge valuable
discussions with Ka Hei Leung, and the contributions of Sergio
Gonzales in the early stages of this work.  This work was supported in
part by the Director, Office of Science, Office of High Energy and
Nuclear Physics, Division of High Energy Physics, of the
U.S. Department of Energy under Contract DE-AC02-05CH11231, in part by
the National Science Foundation under grant PHY-1316783, and in part
by the European Union’s Horizon 2020 research and innovation programme
under the Marie Skłodowska-Curie grant agreements No 690575 and No
674896.

\newpage
\appendix

\def\ksubsection{\Alph{subsection}}
\def\theequation{\ksubsection.\arabic{equation}} 

     
\catcode`\@=11

\def\thesubsection{\Alph{subsection}}
\def\thesubsubsection{\Alph{subsection}.\arabic{subsubsection}}
\noindent{\large \bf Appendix}

\subsection{Cancellation of ultraviolet divergences and evaluation
of the anomaly}\label{uv} 
\noindent
In this appendix, we will specify a choice of PV fields that cancel
leftover divergences from the invariant mass PV sector of~\cite{bg}
(referred to in what follows as BG) and reproduce the universal chiral
anomaly of~\cite{ss}. Aside from the residual divergences discussed
in \myapp{dterms}, the PV fields introduced in BG eliminate all the
divergences from the light sector of the two string models we are
considering, but have some leftover divergences arising from the
$\dot{Y}$ fields if one excludes the noninvariant mass PV sector of
BG. Since we must alter the noninvariant mass sector of BG to produce
a universal anomaly, our strategy here will be to introduce fields
with parameters that decouple as much as possible from the BG fields
but still cancel the divergences of the $\dot{Y}$.  These
new fields replace the BG set that was collectively denoted by $\Psi$.
We expand the sum rules of BG to accommodate the more general K\"ahler
potential of PV fields we consider in \myref{deffh}, and find
that the sum rules that the PV fields must satisfy to cancel
divergences are 

\scriptsize 
\bea \sum_C \eta^C &=& N^\prime
= -N-29\label{a1}\\ 
\sum_\gamma \eta^V_\gamma &=& N_G^\prime = -12 -N_G\\ \sum_C
\eta^C \alpha^C &=& -10\\ \sum_C \eta^C \left(\beta^Cg + \sum_n q^C_n
g^n \right) &=& -A_1 \label{L1}g\\
		\sum_C\eta^C \alpha^C \alpha^C &=& -4\label{a5}\\
		\sum_C \eta^C \delta^C &=& 2\\
		\sum_C \eta^C \left(\alpha^C\beta^C (X_\alpha g_\beta +g_\alpha X_\beta) + \sum_n\alpha^Cq^C_n (g^n_\alpha X_\beta + X_\alpha g^n_\beta )\right) &=& 0\eea
\beq 		\sum_C \eta^C \left(\beta^C\beta^Cg_\alpha g_\beta + \sum_n\beta^Cq^C_n (g^n_\alpha g_\beta + g_\alpha g^n_\beta) + \sum_{n,m}q^C_nq^C_mg^n_\alpha g^m_\beta\right)= -A_2g_\alpha g_\beta - B_2\sum_ng^n_\alpha g_\beta^n \label{L2}\eeq
\bea		\sum_C \eta^C C^C_{(G)} &=& C_{(G)}^M\label{casc}\\
		\sum_C \eta^C (T_a)^C_C\alpha^C &=& 0\\
		\sum_C \eta^C (T_a)^C_C\left(\beta^C+ q^C_n\right) &=& -\sum_p q^p_n (T_a)^p_p\label{L3}\\
		\sum_C\eta^C\left(\frac{F}{2}-\gamma^CF-\sum_n q^C_n F^n\right)\left(\alpha^C-\frac{1}{2}\right)^2X\cdot\tX &=& - \frac{1}{4}\left(\frac{N}{2}-A_1\right)F X\cdot\tX\eea
\noindent
\beq 		\sum_C\eta^C\left(\frac{F}{2}-\gamma^CF-\sum_n q^C_n F^n\right)\left(\alpha^C - \frac{1}{2}\right)\left(2\beta^C X\cdot\tilde{g}+\sum_n 2q^C_n X\cdot\tilde{g}^n\right) = F\sum_p\left(\frac{A_1}{2}-A_2\right)X\cdot\tilde{g}\eeq
\beq 		\sum_C\eta^C\left(\frac{F}{2}-\gamma^CF-\sum_n q^C_n F^n\right)\left(\beta^C\beta^C g\cdot\tilde{g} + \sum_n 2\beta^Cq^C_ng\cdot\tilde{g}^n + \sum_{n,m}q^C_nq^C_m g^n\cdot\tilde{g}^m\right) = -F\left(\frac{A_2}{2} - A_3\right)g\cdot\tilde{g}\label{a14}\eeq
\beq		\sum_C\eta^C\left(\frac{F}{2}-\gamma^CF-\sum_n q^C_n F^n\right) \left(\alpha^C -\frac{1}{2}\right)(T_a)^C_C = \frac{1}{2}
			\sum_p\left(\frac{F}{2} - \sum_nq^p_n	F^n\right)(T_{(G)})^p_p\eeq
\bea		\sum_C\eta^C\left(\frac{F}{2}-\gamma^CF-\sum_n q^C_n F^n\right)(T_{(G)})^C_C \beta^C &=& 0\\ 
		\sum_C\eta^C\left(\frac{F}{2}-\gamma^CF-\sum_n q^C_n F^n\right)(T_{(G)})^C_C q^C_n &=& -\sum_p \left(\frac{F}{2} - \sum_n q^p_n F^n\right)(T_{(G)})^p_pq^p_n\\
		\sum_{C,J}\eta^C\left(\frac{F}{2}-\gamma^CF-\sum_n q^C_n F^n\right)(T_a)^C_J(T_b)^J_C &=&-\sum_{p,q} \left(		
			\frac{F}{2} - \sum_nq^p_n F^n\right)(T_a)^p_q(T_b)^q_p \label{gsg}
\eea
\normalsize
\noindent
where we have used the sum rules for the light field modular weights
\myref{qsums},~\myref{3q} and Eq.~\myref{zero} and
Eq. ~\myref{epsilons}. F is defined by Eq.~\myref{dels}. On the left
hand side of each condition we are summing over all PV fields, while
the right hand sides correspond to summing over the parameters of the
light fields. Since a subset of the the BG PV fields already eliminate
divergences from the light sectors of the string models, we can recast
the above conditions by setting the right hand side of all conditions
to zero and summing over only the $\dot{Y}$, $\Phi$, $\phi$,
$\mathrm{T}$, and $\psi$ fields. To match the anomaly calculated in
SS, we also require
\bea
 0 &=& \sum_C \eta^C(1-2\bar{\gamma}^C)\left(\alpha^C\alpha^C -\frac{2}{3}\bar{\alpha}^C\bar{\alpha}^C - \frac{1}{3}\bar{\alpha}^C	\right)\\
 0 &=& \sum_C \eta^C(1-2\bar{\gamma}^C)\left(2\beta^C\alpha^C - \frac{4}{3}\bar{\beta}^C\bar{\alpha}^C-\frac{1}{3}\bar{\beta}^C\right) \\
 0 &=& \sum_C 	\eta^C(1-2\bar{\gamma}^C)\left(\beta^C\beta^C -\frac{2}{3}\bar{\beta}^C \bar{\beta}^C\right)\\
 0 &=& \sum_C \eta^C \left(1-2\bar{\gamma}^C\right)^2\\
 0  &=&  \sum_C \eta^C q^C_n\alpha^C(1-2\bar{\gamma}^C)\\
 0 &=& \sum_C \eta^C q^C_n\beta^C(1-2\bar{\gamma}^C)\\	
  0 &=& \sum_C \eta^C \bar{\alpha}^C (1-2\bar{\gamma}^C)^2\\
 0 &=&  \sum_C \eta^C \bar{\beta}^C (1-2\bar{\gamma}^C)^2\\
0 &=& \sum_C\eta^C \bar{\alpha}^C (1-2\bar{\gamma}^C)(1-2\bar{\alpha}^C)\\
 0 &=& \sum_C\eta^C\bar{\beta}^C(1-2\bar{\alpha}^C)(1-2\bar{\gamma}^C)\\
 0 &=&  \sum_C \eta^C \bar{\alpha}^C\bar{\beta}^C(1-2\bar{\gamma}^C)\\
 0 &=& \sum_C \eta^C \bar{\beta}^C\bar{\beta}^C(1-2\bar{\gamma}^C)\\
 0 &=&	\sum_C \eta^C \bar{\alpha}^C\bar{\alpha}^C (1-2\bar{\gamma}^C)^2\\
 0&=&	\sum_C \eta^C \bar{\beta}^C\bar{\beta}^C (1-2\bar{\gamma}^C)^2\\
 	0 &=&\sum_C \eta^C \bar{\alpha}^C\bar{\beta}^C (1-2\bar{\gamma}^C)^2\\
 0 &=& \sum_C \eta^C \bar{\beta}^C (1-2\bar{\gamma}^C)\\
 0 &=& \sum_C \eta^C \bar{\beta}^C\bar{\beta}^C\bar{\beta}^C (1-2\bar{\gamma}^C)\\
 30\delta_{nm} &=& \sum_C	\eta^C q^C_n q^C_m(1-2\bar{\gamma}^C)\\
2\sum_p q^p_n +3 -N +N_G &=& \sum_C \eta^C (1-2\alpha^C +2q^C_n)
\eea
\noindent
 In this second set of conditions, only fields with
 noninvariant masses contribute to the sums since $\bar{\gamma}
 =\frac{1}{2}$ for fields with invariant masses.  We now describe a
 particular choice of $\left\{\Phi, \mathrm{T} , \psi\right\}$ fields
 that lead to an easily solvable system for their parameters. We must
 also supplement these fields with the $\phi^C$ fields, since some of
 these have noninvariant masses. Starting with divergences related to
 gauge interactions, we introduce a pair of $\mathrm{T}$ fields for
 each non-Abelian simple factor of the string model gauge group:
 $(\mathrm{T}_{(\G)1}^P,\mathrm{T}_{(\G)1}^{\prime P})$,
 $(\mathrm{T}_{(\G)2}^P,\mathrm{T}_{(\G)2}^{\prime P})$, where $\G$
 specifies the simple group factor. We take the $\mathrm{T}_{(\G)1}$ 
($\mathrm{T}'_{(\G)1}$) to be in the fundamental (antifundamental)
representation of $\G$ while the
 $\mathrm{T}_{(\G)2}$ are gauge singlets. Then Eq.~\myref{casc} gives
\bea  C^M_{\G} &=&  2 C^f_{(\G)}\sum_P \eta_{\mathrm{T}_{(\G)1}}^P \\
 C^f_{(\G)}N_{(\G)} &=& 2N_{\mathrm{T}_{(\G)1}} C^f_{(\G)},\qquad \sum_P\eta_{\mathrm{T}_{(\G)1}}^P = \frac{N_{(\G)}}{2},
 \label{nonab}\eea
\noindent
 for non-Abelian gauge groups and 
\bea
 		 \sum_p Q^p_a Q^p_a = 2\sum_P (\eta_\Phi)^P (Q^\Phi)^P_a (Q^\Phi)^P_a + 2\sum_P (\eta_{\mathrm{T}_{a1}})^P (Q^{\mathrm{T}})^P_a (Q^{\mathrm{T}})^P_a\label{abelcas}
 \eea
\noindent
 for Abelian groups. These are just the conditions needed to cancel
 the $\dot{Y}$ factor of $C^M_{G}$.  The $\Phi$'s enter in
 Eq.~\ref{abelcas} since they are given $U(1)_a$ charges as per the
 prescription in Section 4.2.  Note also that Eq ~\ref{nonab} works
 for the two models considered here since the number $N_{(\G)}$ of
 fundamentals in $\G$ is even for all the gauge groups. We will
 constrain these $\mathrm{T}$ fields so that they do not contribute to
 any divergences other than those arising from gauge interactions. To
 do this, we enforce the following
 \bea
 		(\alpha_{\mathrm{T}_{(\G)2}}) &=& (\alpha_{\mathrm{T}_{(\G)1}})\\ 
		(\beta_{\mathrm{T}_{(\G)2}}) &=& (\beta_{\mathrm{T}_{(\G)1}})\\
		(\alpha_{\mathrm{T}_{(\G)2}}^\prime) &=& (\alpha_{\mathrm{T}_{(\G)1}}^\prime)\label{cond1}\\
		(\beta_{\mathrm{T}_{(\G)2}}^\prime) &=& (\beta_{\mathrm{T}_{(\G)1}}^\prime)\\
		(\eta_{\mathrm{T}_{(\G)2}})^P &=& -(\eta_{\mathrm{T}_{(\G)1}})^P
 \eea
\noindent
  For the $\mathrm{T}$ fields charged under non-Abelian group factors,
  we impose that $\bar{\gamma}^P_{\mathrm{T}_{(\G)1}}$ is independent
  of $P$ so that we can cancel all remaining divergences from
  non-Abelian interactions by demanding
\bea
\frac{C_{\rm GS} - C_{(\G)}}{2} =
 \frac{C^f_{(\G)}N_{(\G)}}{2}\left(1-2\bar{\gamma}_{\mathrm{T}_{(\G)1}}\right),
 \eea
\noindent
 for every non-Abelian group factor $\G$, where 
\beq C_{\rm GS} = 8\pi^2 b = 30\eeq
\noindent
is the adjoint Casimir for $E_8$, which is the gauge group of the pure
Yang-Mills hidden sector of the models considered here.  For the
Abelian divergences, we will not force
$\bar{\gamma}^P_{\mathrm{T}_{(\G)1}}$ to be independent of $P$, but we
will require Eq.~\myref{cond1} and that the primed and unprimed
parameters are identical:
 \bea (\alpha_{\mathrm{T}_{(\G)1}})^P &=& 
(\alpha_{\mathrm{T}_{(\G)1}}^\prime)^{ P}\\
 (\beta_{\mathrm{T}_{(\G)1}})^P &=& 
(\beta_{\mathrm{T}_{(\G)1}}^\prime)^{ P}.\nonumber
 \eea
\noindent
  With these conditions, the remaining divergences from Abelian interactions are canelled by imposing
 \beq  	\frac{C_{GS}}{2} = 
\sum_M (\eta_{\mathrm{T}_{(a)1}})^P \left(1-2\gamma^P_{\mathrm{T}_{(a)1}}\right)(Q^{\mathrm{T}})^P_a(Q^{\mathrm{T}})^P_a.
 \eeq
\noindent
 With the above restrictions, the charged $\mathrm{T}$ fields will
 eliminate only gauge-related divergences and not contribute to any of
 ther other sum rules listed above.  Turning to the $\psi$ and $\phi$
 fields, we choose parameters such that
 \bea
 		\bar{\gamma}_\psi^P = \bar{\alpha}^P_\psi = \bar{\beta}^P_\psi = 0.
 \eea
\noindent
 Then the second set of conditions reduces to 
 \bea
 	0 &=& \sum_P \left(\eta_\psi\right)^P + \sum_C \hat{\eta}^C\left(1-2\bar{\hat{\alpha}}^C \right)^2\\
 	0 &=& \sum_P \left(\eta_\psi\right)^P\left(\alpha_\psi\right)^P\left(\alpha_\psi\right)^P\ddd
 + \sum_C \hat{\eta}^C(1-2\bar{\hat{\alpha}}^C)\left(\hat{\alpha}^C\hat{\alpha}^C - \frac{2}{3}\bar{\hat{\alpha}}^C\bar{\hat{\alpha}}^C -\frac{1}{3}\bar{\hat{\alpha}}^C\right)\\ 
 	0 &=& \sum_C \hat{\eta}^C \bar{\hat{\alpha}}^C \left(1-2\bar{\hat{\alpha}}^C \right)^2 \\
 	0 &=& \sum_C \hat{\eta}^C \bar{\hat{\alpha}}^C\bar{\hat{\alpha}}^C \left(1-2\bar{\hat{\alpha}}^C \right)^2 \\
 	2\sum_p q^p_n +3 -N +N_G &=& \sum_C \hat{\eta}^C (1-2\hat{\alpha}^C)\\
 	0 &=& \sum_P (\eta_\psi)^P\\
	0 &=&\sum_P (\eta_\psi)^P (\alpha_\psi)^P(\alpha_\psi)^P\\
	0 &=& \sum_P (\eta_\psi)^P (\alpha_\psi)^P(\beta_\psi)^P\\
	0 &=& \sum_P (\eta_\psi)^P (\beta_\psi)^P(\beta_\psi)^P\\
	0 &=& \sum_P (\eta_\psi)^P q^P_\psi \alpha_\psi^P\\
	0 &=& \sum_P (\eta_\psi)^P q^P_\psi \beta_\psi^P\\
	30 &=& 2\sum_P (\eta_\psi)^P q^P_\psi q_\psi^P
 \eea
\noindent
 Finally, to cancel all the divergences as required by the sum rules
 in~\myref{a1}--\myref{a14}, we introduce gauge singlet $\mathrm{T}$
 fields $(\mathrm{T}_{3}^P,\mathrm{T}_{3}^{\prime P})$ with invariant
 masses. These fields, along with the $\Phi$ and $\psi$ fields, are
 enough to regulate those divergences of the $\dot{Y}$ fields that do
 not involve gauge couplings. While we have solved this system to obtain
 numerical solutions, the results are not particularly enlightening
 and we will not reproduce them here.

\setcounter{equation}{0}
\subsection{Residual linear and logarithmic divergences}\label{dterms}
There are two sources of the chiral anomaly involving space-time
curvature.  The first arises from the spin connection in the fermion
covariant derivatives.  The three sum rules in \myref{sums}
assure that the linear divergent terms from the PV fermion spin
connection cancel those from the light fields, and the residual
anomaly arises from the PV masses, giving a supersymmetric contribution
\beq  \Del\L_{\rm spin} =
-b_{\rm sp}\int E F\Omega_{\rm GB}\hc,
\label{spin}\eeq
\noindent
which is the variation of the first term in $\L_0$ in \myref{totomega}, 
with
\beq 8\pi^2b_{\rm sp} = {1\over24}\(N' - N'_G - 2\alpha + 2\sum_p q^n_p\)
= {1\over24}\(2\sum_p q^p_n + 3 - N + N_G\) = 31\;\; \forall\;\; n
\label{spinan}\eeq
\noindent
for $Z_3$ and $Z_7$ orbifolds without Wilson lines.  The second
contribution arises from the affine connection in the gravitino
covariant derivative; it has no PV counterpart and is not canceled.
However there is a residual conformal anomaly associated with the
linear divergence arising from the Gauss-Bonnet term which is a total
derivative, and which is uniquely determined~\cite{GB} by the spins of
the particles in the loop. For PV regulated supergravity we have
\beq \L_{\rm G B} =
{\sqrt{g}b_{\rm B G}\over2}\(r_{\mu\nu\rho\sigma}r^{\mu\nu\rho\sigma} -
4r_{\mu\nu}r^{\mu\nu} + r^2\)\ln\Lambda,\label{defLGB}\eeq
\noindent
with 
\beq 8\pi^2b_{\rm G B} = {1\over48}(N + N' - 3N_G - 3N'_G + 41) = 1.
\label{bGB}\eeq
\noindent
The variation of \myref{defLGB} forms a supersymmetric operator with
the chiral anomaly from the gravitino affine connection provided the cut-off
takes the value in \myref{lambda}, giving a contribution 
\beq \Del\L_{\rm aff} = b_{\rm G B}\int E F\Omega_{\rm GB}\hc,
\label{GBan}\eeq
\noindent
which is the variation of the $K\Omega_{\rm G B}$ term in
\myref{totomega}, and combines with \myref{spinan} to give
\beq \Del\L_{\rm anom} \ni - b\int E F\Omega_{\rm GB}\hc,
\label{totGB}\eeq
where
\beq 8\pi^2b = 8\pi^2(b_{\rm sp} - b_{\rm G B}) = {1\over24}\(2\sum_p
q^p_n - N + N_G - 21\) = 30.\qquad \forall\;\; n\label{univ2}\eeq
\noindent

There is also a linear divergence arising from an off-diagonal
gravitino-gaugino connection in the fermion covariant derivative.
This also combines with an uncanceled logarithmically divergent total
derivative to form an anomaly supermultiplet if the cut-off satisfies
\myref{lambda}. It was shown in Appendix B.3 of~\cite{bg} that this
anomaly can be canceled for a particular choice of masses for certain
PV fields that regulate gauge and gravity sector loops.

Finally, there are ``D-term'' anomalies that arise from uncanceled
logarithmically divergent terms with no chiral anomaly counterpart.
These were simply dropped in the evaluation of on-shell ultraviolet
divergences in~\cite{gj} and~\cite{gjs}.  Since they have no chiral
anomaly partner they are more difficult to identify than the above
terms. With the cut-off \myref{lambda}, the conformal anomaly includes
a contribution
\beq \Del\L_{\rm conf} \ni  {\sqrt{g}\over8\pi^2}
\re F K_{i\m}D^\mu\[\D_\mu\z^{\m}\(\l 2F^i\r\l R\r - D_\nu\D^\nu z^i\) +
D_\nu\D_\mu\z^{\m}\D^\nu z^i\hc\],\label{chiganom}\eeq
\noindent
where
\beq F^i = - {1\over4}\D^2Z^i \label{defF}\eeq
\noindent
is the auxiliary field of the chiral supermultiplet $Z^i$, that was
identified in~\cite{bg} as arising from a total derivative dropped in
the evaluation~\cite{gj} of UV divergences for gravity coupled to
chiral matter. When Yang-Mills couplings are included~\cite{gjs},
there are many more terms, and digging out total derivatives is much
more difficult.  We find the additional light field
contribution:
\beq \Del\L_{\rm conf} \ni - \l{\sqrt{g}\over8\pi^2}\re F D^\mu\[
{3\over2}K_{i\m}\D_\mu\z^{\m}D^a(T_a z)^i + {\pp_\mu\s\over s + \s}
D^a D_a\]\r\hc,\label{gjsanom}\eeq
\noindent
where
\beq D^a = - \half\Da\Wa^a\label{defD}\eeq
\noindent 
is the auxiliary field for the superfield strength $\Wa^a$, and we
evaluated the result of \cite{gjs} using the classical K\"ahler
potential in \myref{kdil} for the dilaton.  We also find a
contribution~\cite{pv} from the PV sector
\beq \Del\L_{\rm conf} \ni  \l{\sqrt{g}\over16\pi^2}\re F D^\mu\[
K_{i\m}\D_\mu\z^{\m}D^a(T_a z)^i - {\pp_\mu\s\over s + \s}\(F^i\bF_i
- 8R\bR\)\]\r\hc\label{pvdanom}\eeq
\noindent
With the classical K\"ahler potential in \myref{kdil} the equations of
motion give 
\beq \l F^s\r = \l 2(s + \s)\bR\r, \qquad
\l \bF_s\r = \l{2\over s + \s}R\r, \label{eom}\eeq
\noindent
and the dilaton-dependent contribution can be written
\bea \Del\L_{\rm conf}(s,\s) &=& - {\sqrt{g}\over8\pi^2}
\re F D^\mu\[\(\pp_\mu\s\Box s - \pp_\nu\pp_\mu\s\pp^\nu s\hc\)\right.
\ddd \l
+ \half{\pp_\mu\s\over s + \s}\l\(F^p\bF_p - 12R\bR + 2D^a D_a\)\r\],
\label{dildterm}\eea
\noindent
However we cannot be certain that we have identified all the
uncanceled total derivatives.  It is also possible that one might be
able to modify the PV sector parameter such that the dilaton
dependence can be canceled, as was the case for F-term anomaly
arising from the off-diagonal gaugino-gravitino connection mentioned
above.


\begin{thebibliography}{99}
\bibitem{bg} D.~Butter and M.~K.~Gaillard,
  Phys.\ Rev.\ D {\bf 91}, no. 2, 025015 (2015),
  Phys.\ Lett.\  B {\bf 679}, 519 (2009).
%
\bibitem{pv} M.~K. Gaillard, {Phys.Lett.} B {\bf 342}, 125 (1995), ibid.
B {\bf 347}, 284 (1995); {Phys. Rev.} D {\bf 58}, 105027 (1998), ibid.
D {\bf 61}, 084028 (2000). 
\bibitem{prev} I. Antoniadis, K.S. Narain and T.R. Taylor, {Phys Lett.}
B {\bf 267}, 37 (1991);
 L. Dixon, V. Kaplunovsky and J. Louis, {Nucl. Phys.}
B {\bf 355}, 649 (1991);
 J.-P. Derendinger, S. Ferrara, C. Kounnas and F. 
Zwirner, {Phys. Lett.} B {\bf 271}, 307 (1991);
 G.L. Cardoso and B.A. Ovrut, {Nucl. Phys.} 
B {\bf 369}, 351 (1992);
 M.K. Gaillard and T.R. Taylor, 
{Nucl. Phys.} B {\bf 381}, 577 (1992);
I.~Antoniadis, E.~Gava and K.~S.~Narain, Phys.\ Lett.\ B {\bf 283}, 
209 (1992) and Nucl.\ Phys.\  B {\bf 383}, 93 (1992);
 G.L. Cardoso and B.A. Ovrut, {Nucl. Phys.} 
{\bf 392}, 315 (1993), {ibid.} B {\bf 418}, 535 (1993);
 G. Girardi and R.  Grimm, {Annals Phys.} {\bf 272},
49 (1999).
\bibitem{ss} C. A. Scrucca and M. Serone, {JHEP}  {\bf 0102},













  019 (2001).
\bibitem{linear} P. Bin\'etruy, G. Girardi, R. Grimm and M. M\"uller,
{Phys. Lett.} B {\bf 265} 111 (1991); P. Adamietz, P. Bin\'etruy,
G. Girardi and R. Grimm, {Nucl. Phys.}  B {\bf 401}, 257 (1993).
%
\bibitem{kahler} G.L. Cardoso and B.A. Ovrut, {Nucl. Phys.} 
{\bf 392}, 315 (1993); 
D.Z. Freedman and B. Kors, JHEP {\bf 0611}, 067 (2006). 
%
\bibitem{GB} I.~Antoniadis, E.~Gava and K.~S.~Narain,
  Phys.\ Lett.\ B {\bf 283}, 209 (1992) and 
Nucl.\ Phys.\ B {\bf 383}, 93 (1992).   
%
\bibitem{gj} M.K. Gaillard and V. Jain, {Phys. Rev.} D {\bf 49}, 1951 
(1994).
\bibitem{gjs} M.K. Gaillard, V. Jain and K. Saririan, 
{Phys. Rev.} D {\bf 55}, 833 (1997).

\end{thebibliography}
\end{document}